\documentclass[11pt,twoside,a4paper]{article}
\usepackage{epsfig}
\usepackage{graphicx}
\usepackage{srt_style}
\def\scr{\scriptsize}  

\begin{document}
\baselineskip .5cm
%
\title{NGC\,1052 -- A study of the pc-scale twin jet}

\author{M.\,Kadler, E.\,Ros, J.\,A.\,Zensus, A.\,P.\,Lobanov and H.\,Falcke}

\affil{Max-Planck-Institut f\"ur Radioastronomie,
              Auf dem H\"ugel\,69, D-53121 Bonn, Germany, mkadler@mpifr-bonn.mpg.de}

%
\begin{abstract}
We present results of a VLBA multi-frequency study of the pc-scale
twin jet in NGC\,1052. We observed this object at epoch 1998.99 with the VLBA
at 5, 8.4, 22 and 43\,GHz both in total and linearly polarized intensity.
The spectral analysis confirms the necessity of a free-free absorbing medium,
obscuring the innermost part of both jets. At 5\,GHz
we found a compact linearly polarized emission region at the base of the 
eastern jet with a degree of polarization of 1.5\,\%. At higher frequencies
there is no evidence for polarization in our data. 
A core shift analysis 
constrains the position of the central engine to $\sim$ 0.03\,pc. The shift
rates of the apparent core position with frequency confirm the strong influence of free-free
absorption in conjunction with steep pressure gradients at the bases of both
jets.
\end{abstract}

\paragraph{\bf Introduction:} 

NGC\,1052 is a nearby elliptical galaxy that
harbors an active galactic nucleus in its center. Its radio
structure ranges from $\sim 20^{\prime\prime}$ down to the smallest
resolvable scales below 1 milliarcsecond. Its bright compact radio core
and its small distance of $D=22.6$\,Mpc 
makes NGC\,1052 a prime target for VLBI studies
of the environment of the central engine believed to be a supermassive
black hole of about $10^{8\pm1}M_{\odot}$.\\
On VLBI scales NGC\,1052
shows a jet/counter-jet structure oriented at a position angle of about
65$^\circ$. Kinematical studies at 15\,GHz
\cite{Ver02} show that the two jets are oriented close to the
plane of the sky.
The distribution of water maser emission in NGC\,1052
on pc-scales was imaged with VLBI by \cite{Cla98}. The maser emission is concentrated in two groups of
unresolved maser spots, lying along the western jet.  Strong
free-free absorption at the corresponding region of the source was found
by \cite{Kel99} (see also \cite{Kam01}). 
The presence of atomic gas at the center of NGC\,1052 is 
evident from HI absorption observations \cite{Kel99} and the X-ray spectrum
of the nucleus \cite{Wea99}.

\paragraph{\bf Imaging and model fitting:}

\begin{figure}[t!]
\centering
\includegraphics[scale=0.66]{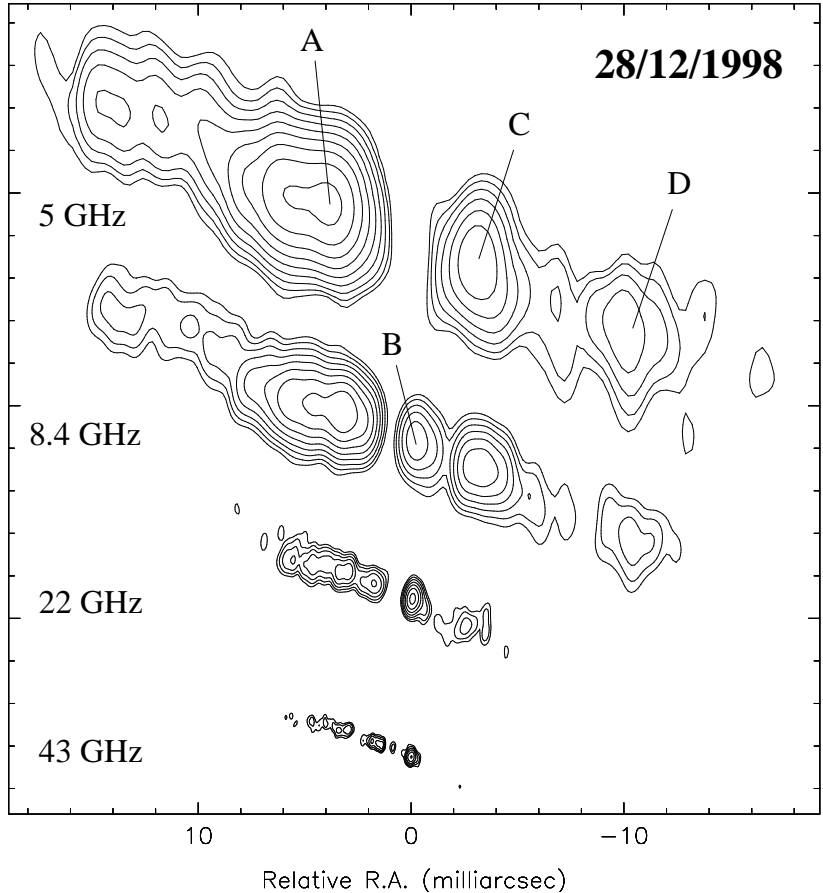}
\caption{VLBA images of NGC\,1052 at 5, 8.4, 22 and 43\,GHz. The map parameters
are given in Table \ref{tab:4maps}.}
\label{fig:aligned}
\end{figure}

\begin{figure}[tbh]
\centering
\includegraphics[scale=0.45, angle=-90]{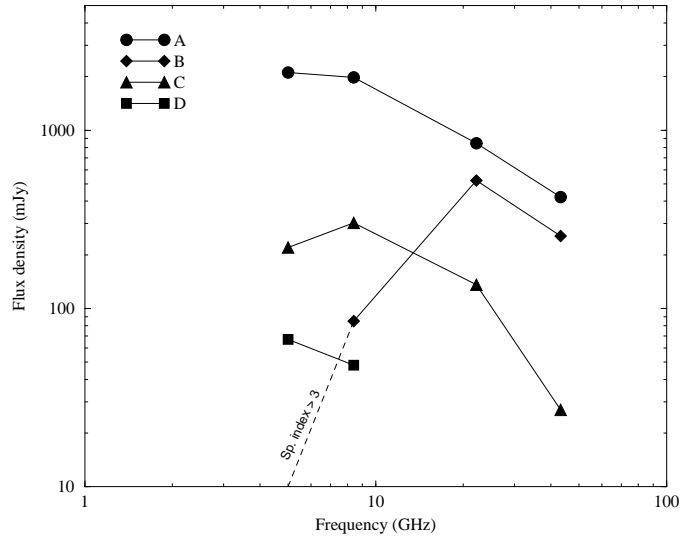}
\caption{Spectra of the four source regions (A, B, C and D in Fig.~\ref{fig:aligned}), obtained from the model fitting of the visibility data.}
\label{fig:spectrum}
\end{figure}
In Fig.~\ref{fig:aligned} we present VLBA images of NGC\,1052 in total 
intensity at 5, 8.4, 22 and 43\,GHz obtained from our observations on 
December 28th 1998
(related parameters in Table~\ref{tab:4maps}). The
images have been produced following standard procedures.
In the images we define four regions, A (eastern jet), B, C and D (western
jet). The alignment of the images at the four frequencies was performed 
through a pairwise identification of Gaussian model fit components between 
adjacent frequencies. We derived spectra (shown in Fig.~\ref{fig:spectrum})
of the different regions of jet and
counter-jet separately, by adding up the flux densities of the
corresponding model components.\\
\begin{table}[hb]
\begin{center}
\caption{Map parameters for Fig.~\ref{fig:aligned}.}
\label{tab:4maps}
\begin{tabular}{cccccccc}
\hline
\scr  $\nu$ &\scr   $\lambda$   &\scr   beam  &\scr   $S_{\rm peak}$  &\scr   $S_{\rm tot}$$^{\rm (a)}$ &\scr   rms &\scr   Contours \\

\scr  [Hz]  &\scr    [cm]     &\scr  [mas $\times$ mas,$^\circ$] &\scr   [Jy/beam]       &\scr   [Jy]  &\scr   [mJy/beam] &\scr   [mJy/beam] \\ \hline
\scr 5 & \scr 6  & \scr 3.30 $\times$ 1.31 , $-$3.74 &\scr  0.660 $^{\rm b}$ & \scr 2.41 &\scr  0.26 &\scr  0.99 $\times$ (2,4,...,256,512) \\
\scr 8.4 &\scr  4  &\scr  1.98 $\times$ 0.81 , $-$3.87 &\scr  0.538 $^{\rm b}$ &\scr  2.39 &\scr  0.25 &\scr  0.81 $\times$ (2,4,...,256,512) \\
\scr 22 &\scr  1.3  &\scr  0.86 $\times$ 0.32 , $-$7.63 &\scr  0.339 $^{\rm c}$ &\scr  1.51 &\scr  1.20 &\scr  1.02 $\times$ (2,4,...,128,256) \\
\scr 43 & \scr 0.7 &\scr  0.45 $\times$ 0.16 , $-$7.93 &\scr  0.126 $^{\rm c}$ &\scr  0.67 &\scr  0.67 &\scr  1.89 $\times$ (2,4,...,16,32,64) \\
\hline
\end{tabular}
\end{center}
\scr $^{\rm a}$ Total flux density recovered in the map model; $^{\rm b}$ Corresponds to the A component; $^{\rm c}$ Corresponds to the B component.
\end{table}    
Both the eastern and the western jet have a steep spectrum at the highest
frequencies. At low to intermediate frequencies the eastern jet remains
steep between 22 and 8.4\,GHz and becomes flat between 8.4 and 5\,GHz
whilst the western jet exhibits a turnover of its spectrum below 22\,GHz.
The innermost part of the western jet (region B) has a highly inverted
spectrum. No emission is detectable in this region at 5\,GHz, making the
spectral index $\alpha$ ($S\propto \nu^{+\alpha}$) larger than 3. This
value exceeds the theoretical maximal value of 2.5 for synchrotron
self-absorption (see e.g. \cite{Ryb79}) marking a possible region of external
absorption (in an obscuring torus?) at the base of the western jet.

\paragraph{\bf Polarimetry:}
\begin{figure}[t!]
\centering
\includegraphics[scale=0.26]{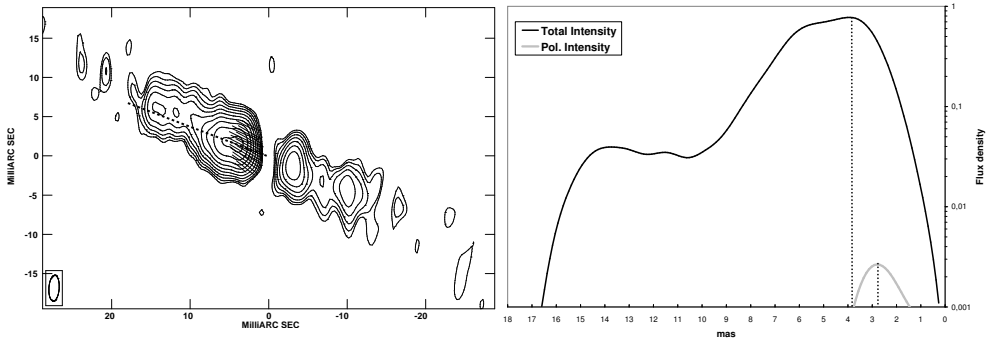}
\caption{Left panel: 5\,GHz image of the total intensity 
(contours) and the linearly polarized emission (segments 
representing the EVPA, proportional to the flux
density). Right panel: profile (dashed line in the left panel) 
of the total and linearly polarized intensity at 5\,GHz taken along
P.A.$=69^\circ$.}
\label{fig:polint}
\end{figure}

We carried out a polarimetric imaging analysis following standard procedures
(see \cite{Lep95}). We detected linearly polarized emission above a
3$\sigma$ level only at 5\,GHz. At this frequency, we found a compact
region of linearly polarized emission at the base of the eastern jet, with
a flux density of about 3\,mJy beam$^{-1}$. This corresponds to a degree
of polarization of 1.5\,\%. The electric field vectors are parallel to the
jet axis, i.e.\ the corresponding emission region is dominated by a
magnetic field perpendicular to the jet axis (assuming no Faraday rotation). 
Our spectral measurements
indicate that the emission from this part of the eastern jet is optically
thick at 5\,GHz. This implies that the polarized flux should increase at
higher frequencies, a trend that we do not observe in our data.
This can be interpreted as an effect of optical depth: if the jet is
linearly polarized only in the outer layers, no linearly polarized
emission is expected at high frequencies where the radiation originates in
deeper layers of the jet, closer to the geometrical jet axis. 

\paragraph{\bf Core-shift analysis:}

\begin{figure}[t!]
\centering
\includegraphics[scale=0.6]{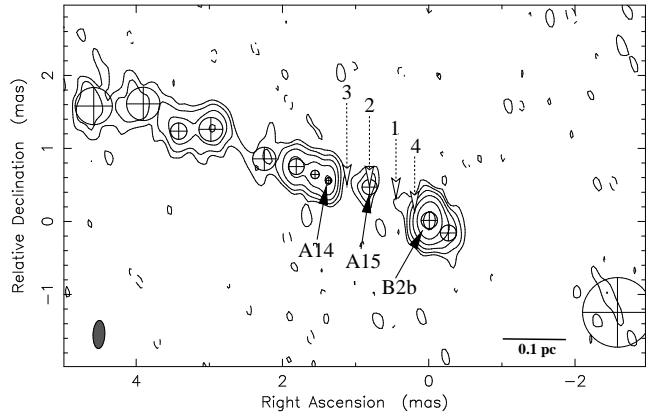}
\caption{Image with the model fitting results of the 
43\,GHz data.  The circles represent the Gaussian components.
The innermost jet components are labeled 
as A\,14, A\,15 and B\,2b. The putative locations of 
the central engine are indicated (dashed arrows) 
for 4 different scenarios (see discussion in the text).}
\label{fig:Qmodel}
\end{figure}

\begin{figure}[t!]
\centering
\includegraphics[scale=0.5,angle=-90]{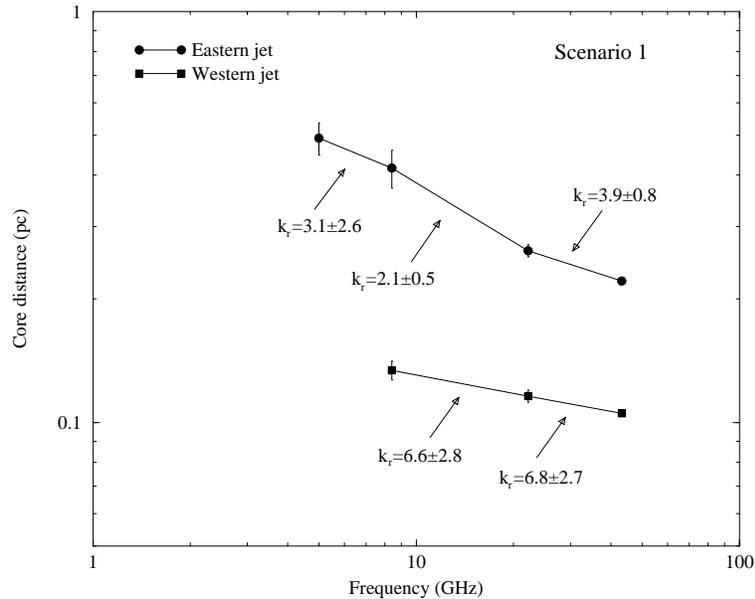}
\caption{Core positions in the two jets at the different 
frequencies for scenario 1 (see Fig.~\ref{fig:Qmodel}). The 
corresponding values of $k_r$ for the core shift between adjacent frequencies
are labeled. Smaller core shifts correspond to higher
values of $k_r$.  Table~\ref{tab:k_r} provides the values of $k_r$ for 
the other three scenarios.}
\label{fig:k_r}
\end{figure}

The symmetry between the jet and the counter-jet constrains the position
of the central engine in NGC\,1052. The location where a jet becomes
visible at a given frequency is usually referred to as the ``core''.
The core is located at the distance to the central engine $r_c$, where
the optical depth $\tau$ has fallen to be $\sim1$. This distance is given
by:  $r_c \sim \nu^{-\frac{1}{k_r}}$,
where $k_r=((3-2\alpha)m+2n-2)/(5-2\alpha)$, with $m$ and $n$ the power
indices of the magnetic field and the particle density: 
$B \sim r^{-m}$, $ N \sim r^{-n}$ \cite{Lob98}. 
Measuring $r_c$ at two frequencies allows one to determine
$k_r$ in the corresponding region of the jet.
For a freely expanding jet in equipartition
\cite{Bla79}, $k_r$=1. The value of $k_r$ is larger in regions with steep pressure
gradients and may reach 2.5, for reasonable values of $m,n$ ($\leq$4, see
\cite{Lob98}).  If external absorption determines the apparent core
position, comparable density gradients of the external medium can alter
$k_r$ to values above 2.5. \\
The values of $k_r$ deduced depend crucially on the absolute values of
$r_c$ on the two sides and therefore on the assumed position of the central
engine. Four scenarios have been tested with different reference points
(see Fig.~\ref{fig:Qmodel}). Table~\ref{tab:k_r} gives the derived values of 
$k_r$ for each scenario. The area between the model components A\,15
and B\,2b is the most likely location of the central engine and the center
between both components is a natural choice for its exact position 
(scenario 1). Shifting the
reference point eastwards (scenarios 2 and 3) alters the values of $k_r$
into unphysically large regimes (requiring density gradients $\propto
r^{-10}$ and higher). Assuming the true center of activity to be located 
more westwards (closer to B\,2b, scenario 4), the values of $k_r$ derived
are still acceptable.
\begin{table}[b!]
\centering
\caption{$k_r$ values for the four different putative centers of 
activity.}
\label{tab:k_r}
\begin{small}
\begin{tabular}{@{}c|c@{~}c|c@{~}c|c@{~}c|c@{~}c@{}}
\noalign{\smallskip}
$\scr \nu$& \multicolumn{2}{c|}{\scr ---Scenario 1---}& \multicolumn{2}{c|}{\scr ---Scenario 2---}& \multicolumn{2}{c|}{\scr ---Scenario 3---}& \multicolumn{2}{c}{\scr ---Scenario 4---} \\
\scr [GHz] &\scr $k_{r,{\rm east}}$&\scr $k_{r,{\rm west}}$\scr & $k_{r,{\rm east}}$&\scr $k_{r,{\rm west}}$ &\scr $k_{r,{\rm east}}$&\scr $k_{r,{\rm west}}$ &\scr  $k_{r,{\rm east}}$& $k_{r,{\rm west}}$ \\
\noalign{\smallskip}
\hline                                                                         
\noalign{\smallskip}
\scr 5--8.4     &\scr 3.1$\pm$2.6 &\scr      --     &\scr 2.5$\pm$2.2 &\scr --           &\scr 3.0$\pm$3.6 &\scr --
&\scr 3.4$\pm$2.8 &\scr --          \\
\scr8.4--22    &\scr 2.1$\pm$0.5 &\scr 6.6$\pm$2.8 &\scr 1.5$\pm$0.3 &\scr 11.9$\pm$5.1 &\scr 1.0$\pm$0.2 &\scr 14.8$\pm$6.2
&\scr 2.4$\pm$0.6 &\scr 3.8$\pm$1.6 \\
\scr 22--43     &\scr 3.9$\pm$0.8 &\scr 6.8$\pm$2.7 &\scr 2.4$\pm$0.5 &\scr 13.0$\pm$5.2 &\scr 1.3$\pm$0.2 &\scr 17.1$\pm$7.1
&\scr 4.8$\pm$1.0 &\scr 3.6$\pm$1.3 \\
\noalign{\smallskip}
\hline
\end{tabular}
\end{small}
\end{table}     
We show the core positions of both jets at the different frequencies for
the first case (scenario 1) in Fig.~\ref{fig:k_r}.  The eastern
jet has rather high values of $k_r$ below 22\,GHz, although still in
agreement with steep pressure gradients in the jet environment. Above
22\,GHz $k_r$ is 3.9$\pm$0.8, which is a good indicator for free-free
absorption affecting the jet opacity. The western jet has values of $k_r$
as high as 6.8$\pm$2.7 between 22 and 43\,GHz, suggesting a large
contribution from free--free absorption. \\
Our results from the core--shift analysis support the picture of a
free--free absorbing torus covering mainly the inner part of the western
jet and also a smaller fraction of the eastern jet. The true center of
activity in NGC\,1052 can be determined to lie between the model components A\,15 and B\,2b, with an uncertainty of only $\sim 0.03$\,pc.

\acknowledgments We want to thank A.\,Tarchi for his valuable help. The VLBA is
operated by the National Radio Astronomy Observatory (NRAO), a facility of the 
National Science Foundation operated under cooperative agreement by Associated
Universities, Inc.

%

%
%
\end{document}